# Direct Observation of Plasmon Band Formation and Delocalization in Quasi-Infinite Nanoparticle Chains

Martin Mayer,[†,§] Pavel L. Potapov,[⊥] Darius Pohl,[§,#,∥] Anja Maria Steiner,[†,§] Johannes Schultz,[⊥] Bernd Rellinghaus,[§,#,∥] Axel Lubk,*[⊥] Tobias A. F. König,*[†,§,‡] and Andreas Fery*[†,§,‡]

[†]Institute of Physical Chemistry and Polymer Physics and [‡]Department of Physical Chemistry of Polymeric Materials, Leibniz-Institut für Polymerforschung Dresden e.V., Hohe Strasse 6, 01069 Dresden, Germany

[§]Cluster of Excellence Center for Advancing Electronics Dresden (cfaed) and [∥]Dresden Center for Nanoanalysis, Technische Universität Dresden, D-01062 Dresden, Germany

[⊥]Institute for Solid State Research and [#]Institute for Metallic Materials, Leibniz-Institut für Festkörper und Werkstoffforschung, Helmholtzstrasse 20, 01069 Dresden, Germany

Ⓢ Supporting Information

**ABSTRACT:** Chains of metallic nanoparticles sustain strongly confined surface plasmons with relatively low dielectric losses. To exploit these properties in applications, such as waveguides, the fabrication of long chains of low disorder and a thorough understanding of the plasmon-mode properties, such as dispersion relations, are indispensable. Here, we use a wrinkled template for directed self-assembly to assemble chains of gold nanoparticles. With this up-scalable method, chain lengths from two particles (140 nm) to 20 particles (1500 nm) and beyond can be fabricated. Electron energy-loss spectroscopy supported by boundary element simulations, finite-difference time-domain, and a simplified dipole coupling model reveal the evolution of a band of plasmonic waveguide modes from degenerated single-particle modes in detail. In striking difference from plasmonic rod-like structures, the plasmon band is confined in excitation energy, which allows light manipulations below the diffraction limit. The non-degenerated surface plasmon modes show suppressed radiative losses for efficient energy propagation over a distance of 1500 nm.

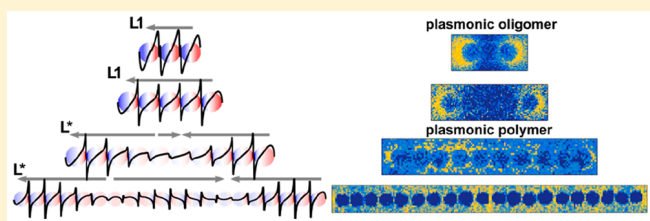

**KEYWORDS:** Surface plasmons, nanoparticle, plasmonic polymer, template-assisted self-assembly, electron-energy loss spectroscopy

Localized surface plasmon resonances (LSPR) are self-sustaining resonances appearing when delocalized conduction-band electrons of a metal are confined within a nanoparticle.[1,2] Plasmonic resonances are characterized by strong and localized electromagnetic field enhancement, which is strongly sensitive to the geometry and composition of the nanoparticle and the environment.[3] This makes them attractive for a wide range of applications, in which sub-wavelength control of electromagnetic fields from the infrared to ultraviolet range is crucial.[4] In particular, long metallic nanoparticle chains have been proposed for plasmonic waveguiding,[5] i.e., photonic transport confined to the submicron and, hence, subwavelength length scale, which is difficult to realize with optical devices.[6,7] In this length scale, plasmonic waveguides open up new strategies for signal transport due to their strong confinement and the high signal speed. It has been predicted through analytical and numerical studies that regular nanoparticle chains support plasmon modes with distinct dispersion relations and, hence, signal transmission velocities depending on the geometric parameters of the chain.[8] More recently, geometric modifications of the monopartite chain prototype, such as bipartite chains or zig-zag chains, are predicted to feature more

complicated band structures, including plasmonic band gaps as well as topological edge states.[9]

For the realization of such waveguides and the experimental proof of the predicted effects, however, energy dissipation in the metal and precise positional control of the metallic nanoparticles are still bottlenecks. In particular, the intrinsic losses of the employed metals limit the overall performance of plasmonic waveguides.[10] However, smart assembly, resulting in finely tuned particle coupling, can further lower the dissipation. For example, Liedl et al. recently showed low-dissipative and ultrafast energy propagation at a bimetallic chain with three 40 nm particles.[11] Various colloidal techniques, such as DNA origami[12] or self-assembly by chemical linkers,[13] are available to assemble noble metal nanoparticles into chains. In principle, infinitely long particle chains can be fabricated with these techniques; however, trade-offs in particle spacing, particle size, or linear geometry must be accepted. In recent years, masks have been used for colloidal self-assembly to overcome the size limitation, which results into long-range energy transfer, even around a micrometer-sized corner.[14] It has been shown that above a rather undefined chain length, the so-called "infinite chain limit", the longitudinal plasmonic modes converge to a nonzero asymptotic energy.[15,16] Consequently, the plasmonic response above this lengths differs from shorter chains because of the discrete nature of the chain and the finite coupling strength.[17] To make it easier to distinguish between them, chains above the infinite chain limit are referred to as plasmonic polymers, whereas shorter chains are called plasmonic oligomers, in close analogy to organic polymer synthesis.[8,12]

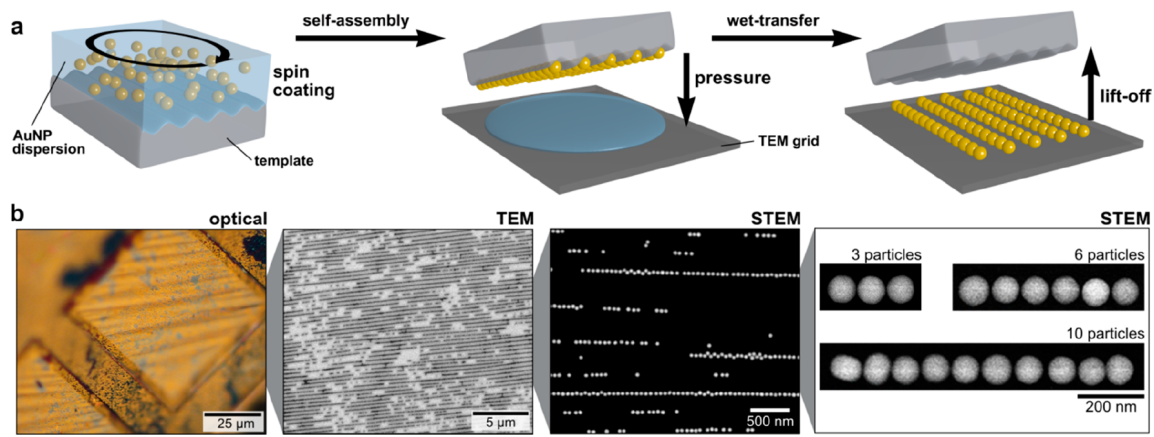

**Figure 1.** Large-area template-assisted self-assembly of various chain lengths and wet transfer for spectroscopic (EELS) studies. (a) Schematic template-assisted colloidal self-assembly via spin coating followed by wet-transfer printing on a TEM grid. (b) A microscope image and (S)TEM images and selective details (3, 6, and 10 particle chains) of the transferred gold particle chains.

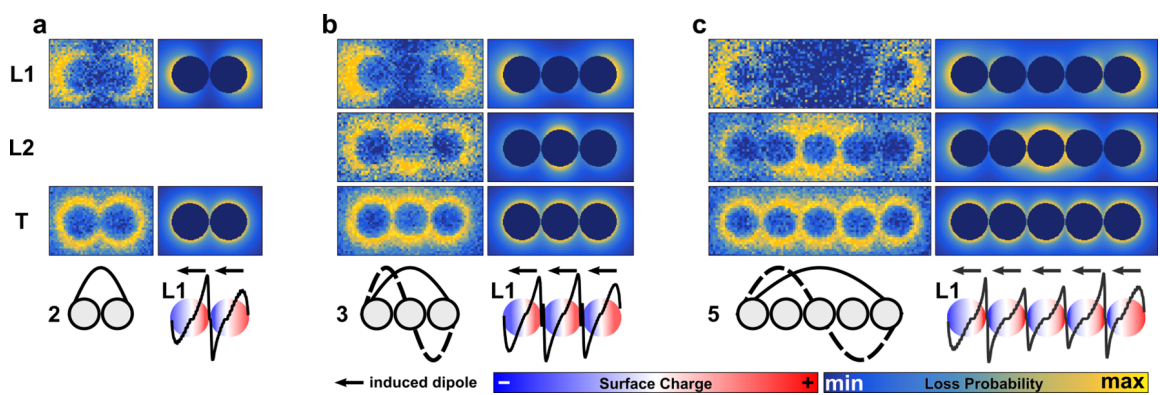

**Figure 2.** Theoretical and experimental plasmonic modes of a (a) dimer, (b) trimer, and (c) pentamer. Schematic descriptions, integrated surface charge images, and plots (black line) as well as corresponding dipole moments for simulated surface charge plots for the most-dominant longitudinal modes. Experimental and simulated EELS maps of longitudinal and transversal modes.

Although the energy transport is principally improved by dark modes, which suffer significantly less from radiation losses than the bright ones, it is still a matter of debate whether those dark modes are responsible for the transport.[18] Recently, methods for visualizing localized plasmons on the nanometer scale have emerged. These methods involve transmission electron microscopy (TEM) combined with electron energy-loss spectroscopy (EELS) with high-energy resolution, which is now readily available in dedicated monochromated TEM instruments. Examples of successful application of TEM and EELS methods includes mapping LSPR modes in metallic nanocubes,[19] nanorods,[20−22] and nanospheres.[23,24]

In this Letter, we fabricated long regular nanoparticle chains, which allow for individual probing of local fields. By comprehensive spatially resolved electron energy loss studies, the plasmonic response is characterized. Here, we will particularly address the transition from individual single particle modes to plasmonic bands in quasi-infinite long chains, which has not been directly observed previously. Robust excitation of the plasmonic waveguide modes relies on recent developments in our groups: single crystalline wet-chemical synthesis[25] and template-assisted colloidal self-assembly[26] as well as improved EELS characterization in the TEM[27] and electromagnetic modeling.[18]

Template-assisted colloidal self-assembly (Figure 1) was used to fabricate colloidal nanoparticle chains (particle diameter of 70 nm) on micrometer-scale carbon-coated TEM grids.[26,28,29] Because high optical quality, reproducibility, and narrow size distributions are crucial parameters, single-crystalline spherical gold nanoparticles (AuNSp) were synthesized by seed-mediated growth, which comply with these requirements.[30,31] The highly linear assembly of these gold spheres into closely packed chains with a homogeneous spacing of <2 nm.[26,31] The obtained interparticle distance relies, in this case, only on the well-defined thickness of the employed dielectric spacer, i.e., in this case, on the protein shell.[26] The resulting spacing can also be seen in the TEM images in Figure 1b. The directed self-assembly process followed by wet-contact printing on the target structure (TEM grid) is outlined in Figure 1a. Transfer to a TEM grid allows the spectroscopic study of the coherent plasmonic coupling in chains of different lengths (Figure 1b) while maintaining the good filling rate (defined by chains per area) and close packing of the nanoparticles within chains. To obtain close-packed chains with varying lengths and a slightly reduced density of chains on the sample, the pH of the colloidal nanoparticle solution was slightly reduced, as described in the experimental section. Thus, such assemblies mark the perfect test system to study the effects of various particle lengths with comparable properties in a combinatorial approach.

The following investigates the transition from plasmonic monomer to polymer (i.e., beyond the infinite chain limit) in detail. Figure 2 shows plasmonic oligomers consisting of a number of nanoparticles well-below the infinite chain limit. The plasmonic properties for such short particle chains with up to five particles have been studied extensively by Mulvaney et al.[24] EELS in the TEM is nowadays a common approach to spatially image plasmons. In this method, the energy loss of a focused electron beam upon crossing the electric field of a plasmon (excited by the evanescent field of the very same electron) is spectroscopically mapped.

For short particle chains, several plasmon modes can be observed in the EELS maps and spectra (see also Figure S4). By directly comparing experimentally observed maps of energy loss with the electromagnetic simulations, the nature of the induced coupling interactions can be elucidated. This comparison allows for the identification of the plasmonic modes by their corresponding surface charges. Starting from a single particle (Figure S2), the addition of a second particle (forming a dimer, Figure 2a) induces hybridization.[32] Hence, the longitudinal modes split into a symmetric (L2) and antisymmetric (L1) one with higher and lower energy, respectively, compared to the fundamental (transversal) mode.[15,24] The maps of energy-loss and the derived surface charges reveal the lower energy antisymmetric mode. The higher-energy L2 mode is not unambiguously detectable because of its lower interaction with the electron beam (see below for details) and its larger damping by interband transitions.

By forming a trimer (Figure 2b), the L1 mode shifts further to lower energies. Additionally, the next order longitudinal L2 mode, with a node in the central particle, can be discerned from the transversal and L1 mode. The mutual cancellation of induced dipoles generates a net dipole moment of zero (even numbers of surface charge waves) rendering this L2 mode dark (i.e., non-radiatively interacting with photons). The higher-energy L3 mode above the transversal mode is again damped by interband transitions. As the number of particles increases to five particles, the energetically lowest mode (L1) approaches the infinite chain limit already (see Figure 2c). However, the induced longitudinal modes L1 and L2 can still be discerned, and the surface charge waves cover the complete length of the chain.

By exciting such particle chains with an electromagnetic field, e.g., a light wave or the evanescent field of a focused electron beam, collective localized surface plasmons are induced in the particle chain. The induced electron density oscillation results in a localized dipole moment of the single particles (defined by surface charges). At large interparticle distances, these oscillations do not couple and are energetically degenerate. When the interparticle distance is decreased far below the excitation wavelength, the localized dipole moments of the single particles couple, lifting the degeneracy. By extension, higher-order multipole moments also couple, which becomes increasingly important at small particle distances. The following shows that the principal plasmonic behavior of nanoparticle chains can be well-described on the dipolar coupling level. In particular the dipoles oriented along the chain axis strongly couple coherently leading to a set of distinguishable longitudinal modes (Figure 3).

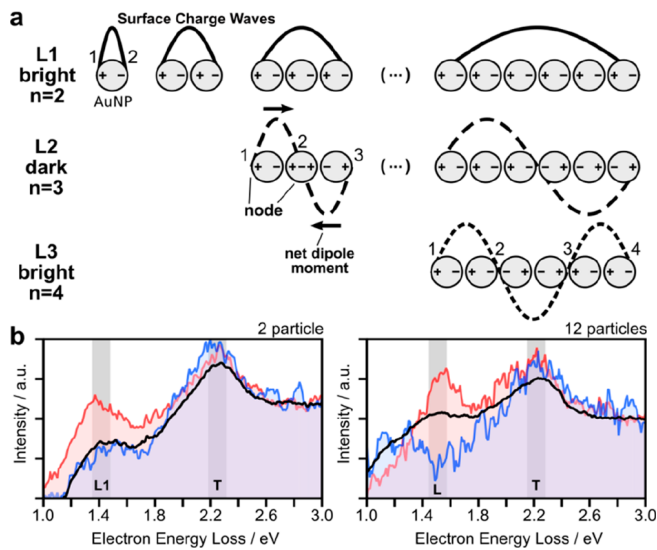

**Figure 3.** Definitions of the longitudinal plasmonic modes and their spectroscopic response. (a) Schematic descriptions of the plasmonic modes along the geometric axis by surface charge (±), dipole moment (black line), net-dipole moment, node, and surface charge wave (schematically representing the polarization field of the respective plasmonic modes). (b) Selected electron energy loss spectra (EELS) for 2 and 12 particles. The blue and red lines show the EELS response averaged over 10–40 pixels, and the black line represents the values averaged over all spectra of the scanned map.

Therefore, similar to the electron-wave function in a diatomic molecule, the energetically lowest coherent plasmonic mode in a plasmonic dimer (L1 at 1.4 eV/885 nm) is an antisymmetric bonding mode (Figure 3b). For longer chains, several harmonics of the longitudinal mode can be excited (see also Figure 3a), which are defined by the number of nodes ($n$). In this context, a node is defined by a zero dipole moment at a specific chain position and the longitudinal modes (L$m$) are indexed by $m = n + 1$.[24] Using this definition, bright modes (nonzero overall dipole moment) occur at odd $m$ and dark modes (zero overall dipole moment) at even $m$, respectively. Figure 3b exemplarily shows two selected EEL spectra for chains with 2 and 12 particles. In contrast to rod like structures,[33] the excitation energy of the L1 mode in long chains converges at low energies, described as the infinite chain limit.[16] In literature, this limit is typically defined somewhere between 8 and 12 particles (see also a spectral visualization in Figure S1).[15,16] In contrast, nanoparticle dipole moments, perpendicular to the chain axis, couple only weakly, resulting in (almost) degenerate transversal modes (marked as $T$ at 2.2 eV/560 nm).

Approaching the infinite chain limit, the intensity of the fundamental longitudinal mode L1 shifts further to smaller energies and gradually vanishes (Figure 4). Electromagnetic simulations reveal that this decrease can be ascribed to an increased damping due to intraband transitions (see Figure S7 for respective spectra). Above L2, the L3 mode becomes observable as identified by the number of nodes ($n = 4$).

When increasing the number of particles in a chain beyond the infinite chain limit, it becomes increasingly difficult or even impossible to discern pure L$m$ modes (Figures 5 and S5 for the maps of 10 and 15 particles) because more modes occupy the same spectral region. Indeed, instead of spectrally well-separated longitudinal modes, distinguished by their node number, a broad band of surface plasmon modes emerges (denoted by L*). This L* mode is characterized by two broad nodes close to the edge and a large maximum in between. The impact of the band formation on the surface charges is summarized in Figure S6. The longest studied chain covers a distance of almost 1.5 μm, although there are much longer chains on the TEM grid (>30 particles) available. However, those particles tend to contain a certain degree of disorder, such as non-colinear alignment, distance variations, etc. These irregularities increasingly influence the mode formation by giving rise to localization effects, as was already visible in Figure 5a (see below for a more-detailed discussion of this effect).

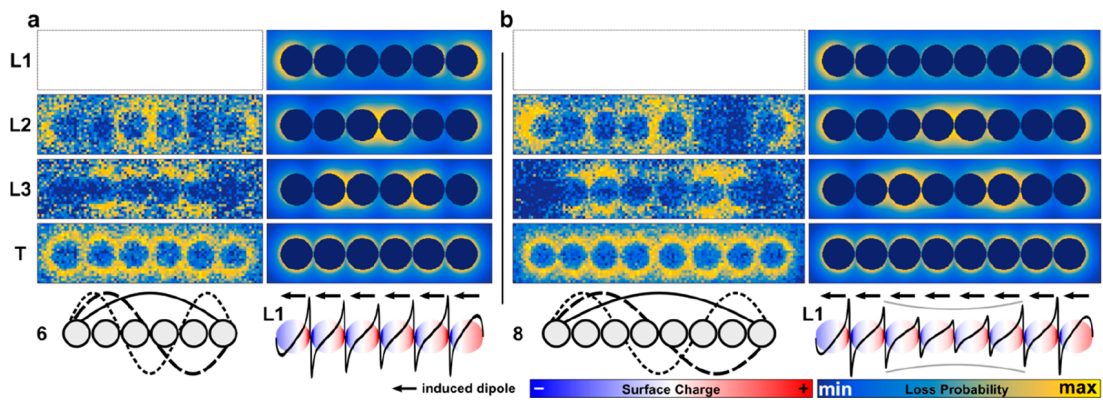

**Figure 4.** Particle chain length approaching the infinite chain limit. Schematic description, simulated surface charge plots of the lowest-order L1 mode, and maps of energy loss for all resolvable plasmonic modes (panels a and b, 6 and 8 particles, respectively; experimental results are shown on the left and simulated results on the right; the intensity of L1 was experientially below the detection limit).

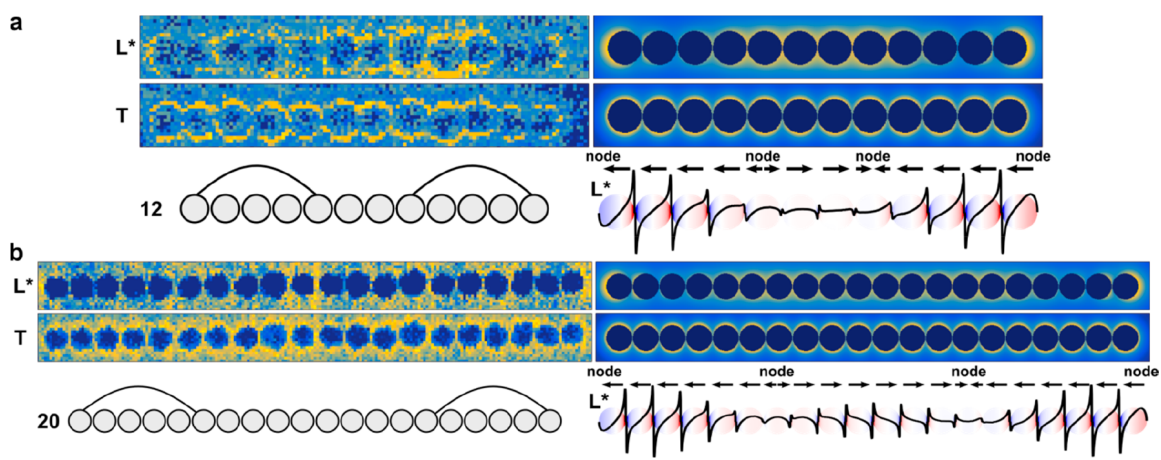

**Figure 5.** Particle chains beyond the infinite chain limit. Schemes, simulated surface charge plots, and maps of energy loss for all selected plasmonic modes (panels a and b show 12 and 20 particles, respectively; experimental results are shown on the left, and simulated results are shown on the right). Note the localized increase of field strength around the small kink of the chain in panel a at the ninth particle of the chain.

Figure 6a summarizes the evolution of the excitation energy upon the transition from plasmonic oligomers to plasmonic polymers by plotting their peak position as a function of the chain length $j$. Significant scattering of the mode energies is observable, which is predominantly due to slight variations (disorder) of the observed chains. For instance, examining the L1 mode of the dimer, the latter consists of "kissing" spheres (Figure 2a), thus leading to an enhanced interaction and, hence, energy shift. The transversal mode is independent from the chain length within the error of the measurement. This indicates small coupling interactions between transversal dipoles and, hence, the degeneracy of the corresponding single particle oscillations. In contrast, more longitudinal modes, separated in energy, appear with increasing chain length. Their energy decreases as the number of particles increases, approaching a lower boundary in the infinite chain limit. Eventually, the longitudinal modes energetically approach each other for plasmonic polymers and superimpose (i.e., degenerate) due to their finite lifetimes and, hence, energy widths (Figure 6b). Consequently, they cannot be discerned anymore (Figure 5), forming an effective L* mode. This degenerated L* mode is defined by its characteristic in-phase excitation at the chain ends and its characteristic superposition of multiple harmonic longitudinal modes in the central region. In the simplified sketch in Figure 6b, the transition from isolated modes to the superimposed L* mode is visualized by an oscillating wave model.

The behavior of long particle chains differs from the reported plasmonic properties of long metallic rods, which support high-order harmonics with well-distinguishable energies.[20,21] To understand the plasmonics of long nanoparticle chains, simple coupled dipole models[8,9] have been used in the past (see Downing et al.[34] for a more elaborate quantum description). Accordingly, nanoparticle chains may be approximated by an assembly of discrete individual dipoles $P_i$ for each particle (of index $i$). These dipoles couple with their neighbors by dipole−dipole interaction facilitated by the electric field propagator $G_{ij}$ (see section S1 of the Supporting Information for an explicit expression of the propagator), as described by the following self-consistent polarization model:

$$P_i(\omega) = \alpha(\omega)\left(E_{ext}(\omega) + \sum_{j \neq i} G_{ij}(\omega)P_j(\omega)\right) \quad (1)$$

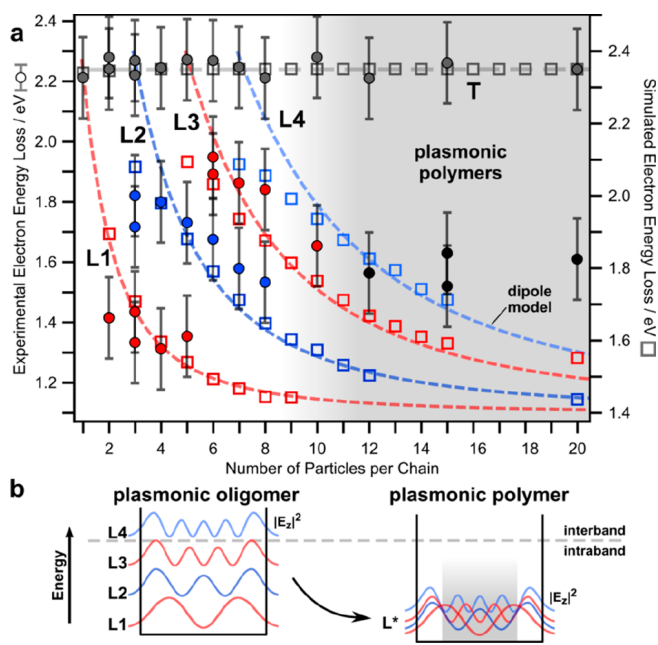

**Figure 6.** Study showing the transition from plasmonic oligomers to plasmonic polymers. (a) Experimental EELS (filled circles) and BEM simulated (hollow squares) energies of the various longitudinal modes as a function of particle length. Dashed lines show the results of the coupled dipole model. (b) Schematic description for the formation of the merged longitudinal plasmonic modes L* beyond the infinite chain limit.

Here, $\alpha$ denotes the (isotropic) single sphere polarizability and $E_{ext}$ the external electric field associated with, e.g., the electron beam. A plasmonic mode occurs at the poles of $(\alpha^{-1} - G)^{-1}$ (with $G \equiv \{G_{ij}\}$), which can be found by searching the zeros of $\det(1 - \alpha(\omega)G(\omega))$. However, $\alpha(\omega)G(\omega)$ is generally non-Hermitian due to retardation and loss (i.e., the complex nature of $\alpha$), and $\alpha$ is a nonlinear function in the complex plane. Therefore, exact zeros (i.e., exact resonances) generally do not exist and maximal responses to external fields occur at the minima of the determinant in the complex plane $\omega$, i.e., at frequencies with imaginary part accounting for the finite lifetime of the plasmon mode. The above model is a simplified version of the more general multiple elastic scattering of multipole expansions method (MESME),[35] which has been previously used to simulate SPR in nanoparticle chains.[36] Indeed, the small interparticle distance also leads to significant quadrupole and even higher-order multipole interactions. Especially, the short distance interaction between nearest neighbors is increased in comparison to the simple dipole interaction model.

Here, the simple structure of the dipole model is exploited to analytically discuss several aspects of the long chain limit, not accessible by a full-scale numerical solution of Maxwell equation. First of all, by evaluating the eigenvalue problem $\overline{G}P(\omega) = a\omega P(\omega)$ (see section S1 in the Supporting Information), eq 1 admits analytical solutions for the mode structure in arbitrary chain lengths, if the interaction to nearest neighbors (i.e., $j = i \pm 1$) is restricted, the inverse polarizability in the considered energy loss regime (i.e., $\alpha^{-1} \approx a\omega$) is linearized, and a $\omega$-averaged propagator (i.e., $G(\omega) \approx \overline{G}$) is used. In particular, the longitudinal mode energies are given by $\omega_l = \omega_0 - 2G_{xx}\cos\left(\frac{l\pi}{n+1}\right)$, $l\in\{1, ..., n\}$, with $n$ denoting the number of particles. The corresponding polarizations read $P_{xj,l} = \sin\left(\frac{lj\pi}{n+1}\right)$, $l\in\{1, ..., n\}$, with $j$ being the particle index. Inserting the dipole field propagator and employing a polarizability derived from the dielectric function of gold,[37] the transverse mode is, however, too high in energy at 2.5 eV, and the energy band is too narrow (0.5 eV) compared to the experiment. To obtain the band widths in reasonable agreement with the experimental results (i.e., approximately that of taking higher-order multipole interactions into account), the coupling parameters were adapted to $G_{xx} \approx 800$ THz (3.3 eV/375 nm) for the dipole aligned along the chain (see Figure 6a). An indirect proof for the existence of higher-order multipole couplings is provided by the full-scale numerical solution of Maxwell's equation, which correctly reproduced the bandwidth. The energy shift can be explained by the interaction with the carbon substrate. The latter acts as a reflecting half-plane for the electric field of the plasmon mode, introducing a second propagator (see section S1 in the Supporting Information for an explicit expression):

$$P_i(\omega) = \alpha(\omega)\left(E_{ext}(\omega) + \sum_{j \neq i} G_{ij}(\omega)P_j(\omega) + \sum_j G_{ij}^{ref}(\omega)P_j(\omega)\right) \quad (2)$$

Primarily, the reflected partial wave directly acts back on the emitting particle, giving rise to an additional self-interaction term $\alpha(\omega)G_{ii}^{ref}(\omega)P_i(\omega)$. This term leads to a renormalization of the polarization and, hence, plasmon energy according to:

$$P_i(\omega) = \frac{\alpha(\omega)}{1 - \alpha(\omega)G_{ii}^{ref}(\omega)}\left(E_{ext}(\omega) + \sum_{j \neq i} G_{ij}(\omega)P_j(\omega)\right) \quad (3)$$

Numerical calculations using the dielectric function of carbon and the geometry of the gold particles show that this effect is sufficient to explain the red shift of approximately 0.2 eV (see also ref 38). In the full-scale solution of the Maxwell equations, an effective medium approach was employed to reproduce the red shift. Considering the various involved approximations, the agreement among full-scale simulations, the dipole model, and the experimental results is rather good (see Figure 6a). The remaining differences to the experimental observed energies are ascribed to long-range coupling of the reflected wave (not included in our models) and deviations from the idealized geometry in the experiment, such as fluctuations in distance, shape, alignment, surface, etc. The latter can be subsumed under disorder effects. While the accurate description of disorder effects requires elaborate perturbation and renormalization schemes (for example, see ref 39 for a treatment within the self-consistent theory of Anderson localization), it is has been established that disorder in 1D systems always tends to localize the wave field with a characteristic exponential damping $|E| \approx \exp(|x - x_0|/\xi(\eta))$ depending on the disorder strength $\eta$.[40] Upon close inspection of the investigated long chains, indeed, the localized excitations are close to geometric perturbations of the chain (e.g., close to the kink in Figure 5a); however, a larger sample base would be required to further elaborate this effect.

Next, the infinite chain limit is discussed using the discrete dipole model. The dipole coupling model describes analytically the formation of a continuous band of longitudinal modes with a dispersion relation (wave vector $q$) $\omega(q) = \omega_0 - 2G_{xx}\cos qd$ in the infinite chain limit ($d$ denotes the interparticle distance). However, due to the finite energy resolution and the finite lifetime (energy width) of the plasmon modes, only a superposition of harmonics can be excited (L* mode). Thus, the highest group velocity $v_{max} = [\partial\omega/\partial q]_{max} \approx 0.4\ c$ is reached around $q = 2\pi/4d$, which corresponds to longitudinal plasmon modes close to the energy of the (nondispersive) transverse mode. The group velocities decrease toward the longer wavelengths (and, therefore, lower excitation energies). Consequently, the group velocity and hence, the transport in the transversal mode is almost zero as the coupling $G_{yy,zz}$ between the transverse dipoles is very small. Finally, the analytical expressions for the net dipole moment are $p_x(q) = \frac{1}{qd}(1 - \cos qL)$ of the $q$-dependent modes (chain length of $L$), which is a measure for the optical coupling strength of individual modes (e.g., optically dark modes have a net moment zero). Accordingly, dark modes appear whenever $q = 2\pi n/L$, and the largest net dipole moment is realized in the long wavelength limit $q \to 0$. Furthermore, from the analytical band dispersion, the plasmonic density of states can be computed according to

$$\text{DOS}(\omega) = 2\frac{dq(\omega)}{d\omega} = \left(G_{xx}d\sqrt{1 - \left(\frac{\omega - \omega_0}{2G_{xx}}\right)^2}\right)^{-1}$$

which also grows toward smaller excitation energies (more generally toward the band edge, which coincides the band minimum in this case). Consequently, we may conclude that the optical coupling ($\sim p \times \text{DOS}$) is maximal for lower excitation energies, where it is eventually bounded by increasing losses.

Thus, treating more complicated chains (e.g., with multipartite basis) including beyond-nearest neighbor interactions requires numerical schemes. However, for periodic arrangements such as the long chain, the use of Bloch's theorem greatly facilitates their implementation (see section S1 in the Supporting Information). In that case, eq 1 reads as;

$$c_{a\mu}(q, \omega) = \sum_{b\nu} D^{b\nu}_{a\mu}(q, \omega)c_{b\nu}(q, \omega) \quad (4)$$

with $c_{a\mu}$ (nanoparticle index in unit cell $a$ and Cartesian index $\mu$) denoting the excitation coefficients of the Bloch waves $P_{a\mu} = c_{a\mu}(q,\omega)e^{iqxn}$ and $D^{b\nu}_{a\mu}$ a dynamical coupling matrix. This formulation has been recently used to compute topological effects in Su−Schrieffer−Heeger model chains.[41]

To highlight the substantial importance of the degenerated L* mode in the plasmonics of linear particle chains, its impact on energy transport along such chains is briefly studied in the following. These considerations will also shine light on an important consequence of the non-Hermiticity (i.e., lossy) nature of the plasmonic system by providing a characteristic length scale for the coherent coupling of the plasmonic modes. Plasmonic waveguiding is chosen because it is typically described as one of the most promising applications for linear particle chains.[11,18,42] Furthermore, the effect is dominated by near-field effects. As it has been suggested earlier, the previously observed degenerated modes have suppressed radiative losses, which may be further tuned by higher multipole order near field coupling terms (beyond dipole).[8,43−45] Finite-difference time-domain (FDTD) electromagnetic modeling were employed to quantify the energy transport properties (Figure 7).

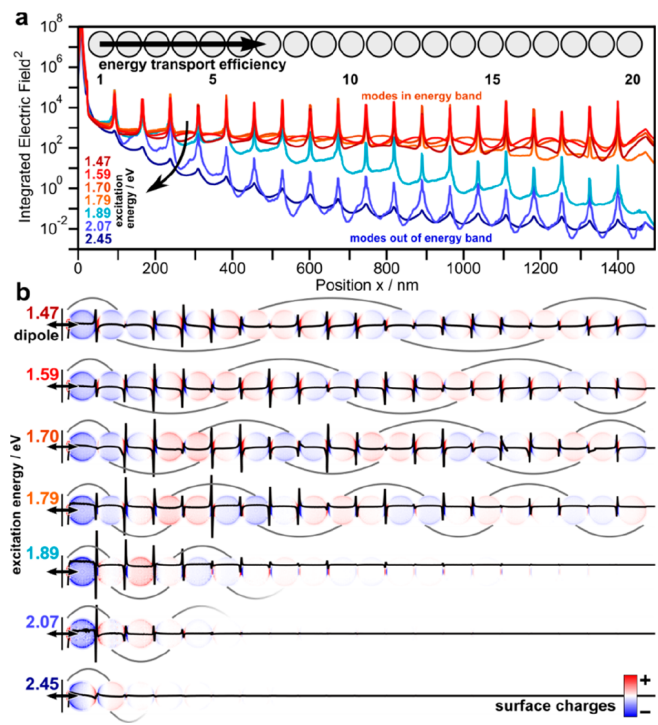

**Figure 7.** Energy-transport properties of plasmonic polymers (consisting of 20 nanoparticles). (a) Integrated electric field along the particle chain in respect to various dipole energies. (b) Corresponding integrated surface charge images (red−blue color scale) and surface charge waves (black line).

As a simplified waveguiding experiment, a dipole source at one end of the particle chain is used as source. Figure 7a visualizes the energy transport efficiency along the chain for all excitable modes by the integrated electric field (red indicating within and blue indicating out of the energy band).

The energetically highest mode (transversal, 2.42 eV, dark blue) exhibits a double exponential decay resulting in a theoretical damping factor of −1.83 dB/50 nm. Longitudinal waveguiding modes above 1.8 eV (i.e., not yet merged with the energy band) perform similar to a fast decay of the transported energy (blue; for the spectral overlap of the individual longitudinal modes, see Figure S8b). On the contrary, degenerated modes overlapping with the energy band support efficient waveguiding with damping factors as low as −0.26 dB/50 nm (for an excitation energy of 1.59 eV). For very small energies, however, the damping factor increases again. This behavior is consistent with previous studies[8] and can be explained by an optimal balance of group velocity (increasing toward higher energies) and radiative losses (decreasing toward high energies) in the sub-radiant level above the super-radiant L1. In comparison to a comparable DNA-assembled waveguide, which is below the infinite chain limit, the herein observed damping is lower by a factor of 3.[18]

Table 1 lists the properties of the supported waveguiding modes as well as the integrated electric fields $E^2$ at the first and last particle for each mode (necessary to calculate the damping factor). The induced surface charge waves (Figure 7b) propagate with an effective energy along the particle chain, as visualized by the respective length of the wave package. The energy propagation for a selected mode (1.59 eV) is highlighted in Video S1. Finally, the mode size perpendicular to the particle chains (energy $E^2$ at $1/e^2$ of the cross-section) is critical for practical applications to avoid cross-talk between neighboring waveguides. The mode size for degenerated modes is below 300 nm, while for the non-degenerated modes it is 140 nm. However, in the real system, disorder effects produce an additional localization (see the discussion above), which may further increase the damping factor.

**Table 1. Plasmonic Waveguide Properties of Different Plasmonic Modes (Transverse Mode: 2.42 eV and Longitudinal Modes: >2.42 eV; Effective Mode Size Describes the Radial Drop of the Integrated $E^2$ perpendicular to the Particle Chains to $1/e^2$)**

| | | | | | | | |
|---|---|---|---|---|---|---|---|
| excitation energy, eV | 2.42 | 2.07 | 1.89 | 1.79 | 1.70 | 1.59 | 1.47 |
| integrated $E_0^2$ ($x_0$ = 0 nm) | 1071 | 1513 | 1925 | 1758 | 1679 | 1519 | 1171 |
| integrated $E_N^2$ ($x_N$ = 1380 nm) | <0.01 | <0.01 | 0.06 | 28 | 218 | 284 | 114 |
| damping in decibels per 1380 nm | −50.4 | −56.2 | −45.0 | −17.9 | −8.8 | −7.2 | −10.1 |
| damping in decibels per 50 nm | −1.83 | −2.04 | −1.63 | −0.65 | −0.32 | −0.26 | −0.37 |
| effective mode energy, eV | 8.86 | 4.43 | 4.13 | 2.82 | 2.58 | 2.07 | 1.70 |
| effective mode size, nm | 137 | 141 | 145 | 173 | 198 | 229 | 261 |

In summary, we directly observed the formation of hybridized plasmonic modes on linear nanoparticle chains as a function of the chain length ranging from one nanoparticle to the infinite chain limit. Plasmonic oligomers and polymers have been fabricated by template-assisted colloidal self-assembly yielding highly ordered chains with low disorder. Electron energy-loss spectroscopy and theoretical modeling allowed us to characterize the plasmonic mode transition between short chains (plasmonic oligomers) and long chains (plasmonic polymers). Plasmonic oligomers show well-separated, non-degenerated longitudinal modes and degenerated transverse modes. Beyond roughly 10 particles, a band of longitudinal modes eventually emerges. This band exhibits a degenerated plasmonic waveguide mode signature with a large dispersion, which makes it suitable for long distance energy transport in the optimal band region. Theoretical simulations of the waveguiding performance suggest that the L* mode with a mode wavelength of 730 nm is able to transport energy with a damping of −0.26 dB/50 nm at a mode cross-section size of 260 nm. Our findings pave the way for further exploitation of plasmonic nanoparticle chains as waveguides and photonic devices. For example, the waveguide mode can be intervened at each contact point by inserting materials such as fluorescent emitters[46] and organic[47] or smart polymers[48] to further modulate and enhance the radiative properties of the plasmonic waveguide.[49] Of particular interest is the coupling at end points of the chain, which sustain particular edge states that may be tuned by complex chain geometries (e.g., bipartite chains). Furthermore, the presented approach opens up new possibilities to study combinatorically the plasmonics of multimetallic chains or chains with multiply particle morphologies within one chain.

**Methods.** *Synthesis, Template-Assisted Colloidal Self-Assembly, and Wet Transfer to Carbon-Coated TEM Grids.* The synthesis of single-crystalline spherical gold nanoparticles with a diameter of 70.5 ± 1.2 nm was accomplished by a three-step seed-mediated growth approach and template-assisted self-assembly into wrinkled PDMS templates via spin-coating of protein (BSA) coated particle solutions (12 mg/mL [Au$^0$]) as published elsewhere.[31] The pH of the nanoparticle solution was adjusted to pH 10 to produce closely packed particle lines in incompletely filled templates.

Transfer in the grooves of the template assembled particle chains was performed by wet-contact printing. The 3 nm carbon-film coated TEM grid (copper, 300 square meshes) was incubated in PEI solution (1 mg/mL, 1800 g/mol, linear) for 1 h and subsequently washed with purified water (Milli-Q-grade, 18.2 MΩ cm at 25 °C). For the transfer, a 2 μL water droplet (pH 10) was placed on the TEM grid, and the particle-filled PDMS stamp was pressed onto the grid with a constant pressure of 100 kPa, and the grid was left to dry under environmental conditions (23 °C, 55% relative humidity). After drying (4 h), the stamp was carefully removed, leaving the nanoparticle chains on the carbon film of the grid.

*STEM EELS Characterization.* STEM scanning and EELS spectrum-imaging was performed in the probe-corrected Titan[3] operating at 300 kV. The microscope was equipped with a Tridiem energy filter and a Wien monochromator operating in the accelerating mode, which ensured the energy resolution of 100−120 meV. EELS was performed under the convergence angle of 22 mrad and the collection angle of 8 mrad with the energy dispersion of 0.01 eV per channel. The spectrum images were acquired with a beam current 200 pA and dwell time 25 ms. For each chain of gold particles, 4−7 runs of spectrum-imaging were performed followed by the summation of the data cubes with accounting for the spatial drift between runs. The obtained spectra were corrected for the energy instabilities, and the zero-loss peak was then removed by using the reference profile collected in a separate run without any sample object. An example of spectra treatment is shown in Figure S3. Finally, the distinct peaks in the low-loss region of spectra were recognized and their energy positions and magnitude were fitted using the nonlinear least-squares procedure. The integrated area under each fitted peak was plotted as a function of the beam position giving rise to maps of the probability for plasmon excitation, i.e., maps of energy-loss.

*Electromagnetic Simulations.* Simulations of electron energy loss spectra and mappings were performed using the Matlab MNPBEM13 toolbox,[50] which is based on the boundary element method (BEM) by García de Abajo.[51] Each sphere of the chain was approximated by triangulation (400 vertices per particle). The dimensions and positions of the gold nanospheres were selected according to the experimental size (70.5 ± 1.2 nm) and spaced with a distance of 1.5 nm. The dielectric properties of gold were taken from Johnson and Christy.[37] An effective medium of $n = 1.2$ is selected to emulate the air−substrate interface. The energy of the simulated electron beam was set to the experimental accelerating voltage of the TEM (300 kV). For each particle chain several spectra were evaluated in the energy range from 1−3 eV at different (not overlapping) electron-beam positions to ensure excitation of all possible plasmonic modes EELS mappings were performed at selected energy levels and were simulated by 1.5 nm meshing of the electron beam. The respective spectra are summarized in Figure S7.

For energy-transfer and waveguiding simulations, a commercial-grade simulator based on the finite-difference time-domain (FDTD) method was used to perform the calculations (Lumerical Inc., Canada, version 8.16).[52] For the broadband and single-mode excitation for energy transport, a dipole source for the specific wavelength range (short pulse length of 3 fs; $E \approx$ 1−4 eV; $\lambda = 300−1300$ nm; see spectra in Figure S8a) and single wavelength to probe individual modes (long pulse length of 18 fs) was used with a distance of 15 nm to the first particle, respectively. To extract the exact peak positions from the spectra, the internal multipeak-fitting function of IGOR Pro 7 (WaveMetrics) was used. The mode size of the modes was defined by integrating of $E^2$ along the particle line and radially defining the $1/e^2$ decay of the $E^2$ field. For the FDTD simulations, the same refractive index data, nanoparticle dimension, inter particle distance, and materials constants were used as in the BEM simulations. Perfectly matched layers in all principal directions as boundary conditions, zero-conformal-variant mesh refinement, and an isotropic mesh overwrite region of 1 nm were used. All simulations reached the auto shut-off level of $10^{-5}$ before reaching 150 fs of simulation time.

## ■ ASSOCIATED CONTENT

### ⑤ Supporting Information

The Supporting Information is available free of charge on the ACS Publications website at DOI: 10.1021/acs.nanolett.9b01031.

Additional simulations, additional EELS maps, experimental EELS spectra including data treatment, and a detailed description of the discrete dipole model (PDF)
A movie showing the electric field during energy propagation along a particle chain (MPG)

## ■ AUTHOR INFORMATION

### Corresponding Authors
*E-mail: a.lubk@ifw-dresden.de.
*E-mail: koenig@ipfdd.de.
*E-mail: fery@ipfdd.de.

### ORCID ●
Martin Mayer: 0000-0003-4013-1892
Tobias A. F. König: 0000-0002-8852-8752
Andreas Fery: 0000-0001-6692-3762

### Notes
The authors declare no competing financial interest.

## ■ ACKNOWLEDGMENTS

This project was financially supported by the Volkswagen Foundation through a Freigeist Fellowship to T.A.F.K. A.L. has received funding from the European Research Council (ERC) under the Horizon 2020 research and innovation program of the European Union (grant agreement no. 715620). P.L.P. acknowledges funding from DFG "Zukunftskonzept" (F-003661-553-Ú6a-1020605).

## ■ REFERENCES